\documentclass[aps,prb,reprint,superscriptaddress]{revtex4-1}

\usepackage{graphicx}
\usepackage{dcolumn}
\usepackage{bm}
\usepackage{courier}
\usepackage{hyperref}
\hypersetup{hypertex=true,
	colorlinks=true,
	linkcolor=blue,
	anchorcolor=blue,
	citecolor=blue}
\begin{document}


\title{Spin excitations in the quantum dipolar magnet Yb(BaBO$_3$)$_3$}


\author{C. Y. Jiang}
\affiliation{State Key Laboratory of Surface Physics, Department of Physics, Fudan University, Shanghai 200438, China}
\author{Y. X. Yang}
\affiliation{State Key Laboratory of Surface Physics, Department of Physics, Fudan University, Shanghai 200438, China}
\author{Y. X. Gao}
\affiliation{School of Physics and Wuhan National High Magnetic Field Center, Huazhong University of Science and Technology, Wuhan 430074, People’s Republic of China}
\author{Z. T. Wan}
\affiliation{School of Physics and Wuhan National High Magnetic Field Center, Huazhong University of Science and Technology, Wuhan 430074, People’s Republic of China}
\author{Z. H. Zhu}
\affiliation{State Key Laboratory of Surface Physics, Department of Physics, Fudan University, Shanghai 200438, China}
\author{T. Shiroka}
\affiliation{Laboratory for Muon-Spin Spectroscopy, Paul Scherrer Institute, CH-5232 Villigen PSI, Switzerland}
\affiliation{Laboratorium für Festkörperphysik, ETH Zürich, CH-8093 Zürich, Switzerland}
\author{C. S. Chen}
\affiliation{State Key Laboratory of Surface Physics, Department of Physics, Fudan University, Shanghai 200438, China}
\author{Q. Wu}
\affiliation{State Key Laboratory of Surface Physics, Department of Physics, Fudan University, Shanghai 200438, China}
\author{X. Li}
\affiliation{State Key Laboratory of Surface Physics, Department of Physics, Fudan University, Shanghai 200438, China}
\affiliation{Key Laboratory of Neutron Physics and Institute of Nuclear Physics and Chemistry, China Academy of Engineering Physics (CAEP), Mianyang 621999, China} 
\author{J. C. Jiao}
\affiliation{State Key Laboratory of Surface Physics, Department of Physics, Fudan University, Shanghai 200438, China}
\author{K. W. Chen}
\affiliation{State Key Laboratory of Surface Physics, Department of Physics, Fudan University, Shanghai 200438, China}
\author{Y. Bao}
\affiliation{Institute of High Energy Physics, Chinese Academy of Sciences, Beijing 100049, China}
\affiliation{Spallation Neutron Source Science Center, Dongguan 523803, China}
\author{Z. M. Tian}
\affiliation{School of Physics and Wuhan National High Magnetic Field Center, Huazhong University of Science and Technology, Wuhan 430074, People’s Republic of China}
\author{L. Shu}
\email{leishu@fudan.edu.cn}
\affiliation{State Key Laboratory of Surface Physics, Department of Physics, Fudan University, Shanghai 200438, China}
\affiliation{Collaborative Innovation Center of Advanced Microstructures, Nanjing 210093, China}
\affiliation{Shanghai Research Center for Quantum Sciences, Shanghai 201315, China}

\date{\today}

\begin{abstract}
We report results of magnetization, specific-heat and muon-spin relaxation measurements on single crystals of disorder-free Yb$^{3+}$ triangular lattice Yb(BaBO$_3$)$_3$. The magnetization experiments show anisotropic magnetic properties with Curie-Weiss temperatures $\theta_{\perp}=-1.40$~K ($H \perp c$) and $\theta_{\parallel}=-1.16$~K ($H \parallel c$)  determined from low temperature data. The absence of both long-range antiferromagnetic order and spin freezing is confirmed down to 0.27 K at zero field. A two-level Schottky anomaly due to the opening of the ground-state Kramers doublet is observed from the low-temperature specific-heat measurements when the applied magnetic fields $\mu_0H >0.7$~T. At zero field, the increase of both $C_{\rm mag}/T$ and the muon spin relaxation rate $\lambda$ below 1~K is due to the electronic spin excitations, which often exist in quantum magnets where dipole-dipole interaction creates an anisotropy of magnetic properties. The spin excitation is also supported by the unusual maximum of field dependence of $\lambda$ due to the field-induced increase of the density of excitations. We argue that dipolar interaction is dominant and induces the spin dynamics in the quantum magnet Yb(BaBO$_3$)$_3$. 
\end{abstract}


\maketitle

\section{\label{sec:level1}INTRODUCTION}
There has been growing interest in quantum magnets which exhibit various exotic phenomena relevant to quantum effects in recent years \cite{keimer2017physics,basov2017towards}. Quantum spin liquid (QSL) is such a novel state in which spins are highly entangled and fluctuate strongly even at zero temperature \cite{anderson1973resonating,savary2016quantum,wen2002quantum,zhou2017quantum}. The QSL state requires the paradigm of Landau's symmetry breaking theory. Instead of order parameters, fractional excitations and the emergent gauge structure are proposed to characterize a QSL state. The search for QSL materials remains an important topic in modern condensed matter physics because of the exotic features above and the relevance to high temperature superconductivity, as well as the potential application in quantum computing \cite{anderson1987resonating,kitaev2003fault}. Geometrically frustrated magnetic systems with antiferromagnetically coupled spins are supposed to host the QSL state \cite{balents2010spin,mendels2016quantum}. Numerous two-dimensional QSL candidates with a frustrated geometry have been reported \cite{shimizu2003spin,itou2008quantum,helton2007spin,nakatsuji2006metallic}. Among them, Yb-based triangular lattice systems have attracted a lot of interest \cite{li2015gapless,ding2019gapless,baenitz2018naybs,dai2021spinon}.\par

Triangular lattices of Yb$^{3+}$ ions with antiferromagnetic interaction show abundant magnetic and quantum features due to the strong quantum fluctuations, providing a promising way to realize QSL candidates \cite{jackeli2009mott,iaconis2018spin,ross2011quantum,li2017spinon,dun2014chemical,ranjith2019field,baenitz2018naybs,ranjith2019anisotropic}. The antiferromagnet YbMgGaO$_4$ with triangular lattice has attracted significant attention due to the possibility of being a gapless U(1) QSL, which is evidenced by many experimental results, including the diffusive spin excitation observed in neutron scattering spectra \cite{shen2016evidence}, power-law temperature dependence of the specific heat \cite{li2015gapless}, persistent spin dynamics at very low temperatures \cite{li2016muon}. However, the ground state of YbMgGaO$_4$ is still controversial due to the unavoidable Mg/Ga disorder, which can lead to a spin-liquid like behavior \cite{zhu2017disorder}. Itamar \textit{et al} proposed that the formation of local singlets can be promoted by magnetic fluctuation under the effect of quenched disorder. The physical properties of YbMgGaO$_4$ at low temperatures are well described by this random-singlet scenario \cite{kimchi2018valence}. To ascertain whether disorder leads to spin-liquid like behavior or QSL survives disorder in YbMgGaO$_4$, Yb triangular lattices without site-mixing disorder are in urgent need.\par

Yb-based triangular lattice single crystals of Yb(BaBO$_3$)$_3$ have been synthesized recently. It was identified as a disorder-free triangular lattice \cite{gao2018synthesis,bag2021realization}. Nuclear magnetic resonance (NMR) study and thermodynamic study on Yb(BaBO$_3$)$_3$ have been performed to figure out its ground state and explore its suitability as a QSL candidate \cite{zeng2020nmr,bag2021realization}. Gapless spin excitations consistent with U(1) QSL ground state with spinon fermion surface were observed~\cite{zeng2020nmr}. However, specific-heat measurements and theoretical calculations suggest that Yb(BaBO$_3$)$_3$ may realize a quantum dipole lattice, where the exchange interactions are negligible~\cite{bag2021realization}. It is worth noting that an energy gap with a gap size proportional to the field intensity was observed both in the NMR and thermodynamic studies. However, the energy gap is believed to be related to the crystal electrical field (CEF)~\cite{bag2021realization}, whereas the NMR study attributes the origin of the gap to spin excitations \cite{zeng2020nmr}.\par

To clarify the origin of the energy gap and provide solid evidence of whether the disorder-free Yb triangular lattice of Yb(BaBO$_3$)$_3$ has a QSL ground state, magnetization, specific heat and $\mu$SR measurements on single crystals of Yb(BaBO$_3$)$_3$ were carried out. At high temperatures, an Orbach process due to the CEF effect determines the muon spin relaxation rates, consistent with the deviation of high-temperature Curie-Weiss behavior around 200~K.  The absence of both long-range antiferromagnetic order and spin-glass behavior is confirmed down to 0.27 K. Only a two-level Schottky anomaly is observed in the magnetic specific heat at low temperatures with magnetic fields applied. At zero field, the extrapolated energy splitting of two-level Schottky anomaly is nearly 0, and increases linearly with magnetic fields. The increase of $C_{\rm mag}/T$ below 1~K at zero field is due to the electronic spin excitations, which often exist in quantum magnets where dipole-dipole interaction is dominant to create an anisotropy. The spin excitation is also supported by the increase of zero field muon spin relaxation rate $\lambda$ below 1~K, and an unusual maximum of field dependence of $\lambda$ due to the field-induced increase of the density of excitations.

\section{EXPERIMENTAL DETAILS}
\begin{figure}[h]
	\centering
	\includegraphics[height=4.0cm,width=8.5cm]{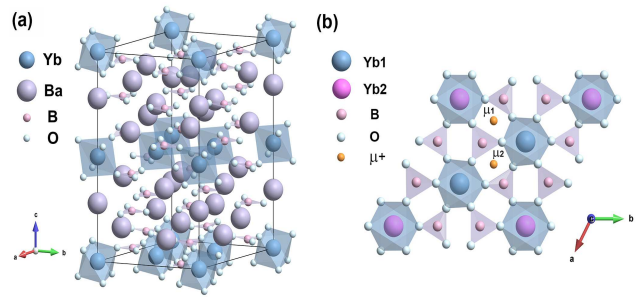}
	\caption{\label{structure}(a) Crystal structure of Yb(BaBO$_3$)$_3$ (space group $P6_3cm$) in one unit cell. (b) The triangular lattice formed by Yb$^{3+}$ cations. The inequivalent Yb$^{3+}$ sites are denoted by Yb1 (blue spheres) and Yb2 (pink spheres). Orange spheres:  candidate $\mu^{+}$ sites $\mu1$ and $\mu2$ (see text).}
\end{figure}

Single crystals of Yb(BaBO$_3$)$_3$ were synthesized by flux method (see supplementary materials in Ref.~\onlinecite{zeng2020nmr}). The sample structure was checked by both single crystal and powder X-ray diffraction (XRD) measurements. The crystal structure is shown in Fig.~\ref{structure}(a). The structural refinement reveals that Yb(BaBO$_3$)$_3$ is free of chemically random mixed occupancies of Yb$^{3+}$ and B$^{3+}$/Ba$^{2+}$ cations \cite{gao2018synthesis,zeng2020nmr}, reflecting the large difference of ionic radii of Yb$^{3+}$ and B$^{3+}$/Ba$^{2+}$ cations in Yb(BaBO$_3$)$_3$. Thus, only fully ordered arrangements of Yb$^{3+}$ triangular lattices are formed (see  Fig. \ref{structure}(b)). Yb$^{3+}$ ions coordinated with 6 nearest neighboring O atoms form YbO$_6$ octahedrons which are linked by corner-shared BO$_3$ triangles. Three layers of non-magnetic ions Ba/B/O are stacked along the $c$ axis between every two layers of Yb$^{3+}$ ions, so that the compound can be treated as a quasi two-dimensional triangular lattice.\par

Magnetization of Yb(BaBO$_3$)$_3$ was measured in a superconducting quantum interference device magnetometer (Quantum Design magnetic property measurement system). The temperature-dependent susceptibility from 2 to 300~K was measured under a magnetic field of 1~T in both zero-field-cooled (ZFC) and field-cooled (FC) procedures. 

Specific heat of Yb(BaBO$_3$)$_3$ down to 0.3~K was measured by the adiabatic relaxation method in a Quantum Design physical property measurement system (PPMS) equipped with a Dilution Refrigerator option. A nonmagnetic polycrystalline sample of Lu(BaBO$_3$)$_3$ was also grown and its specific heat was measured to determine the phonon contribution in the total specific heat of Yb(BaBO$_3$)$_3$. \par 

The muon spin relaxation measurements were carried out at the Dolly spectrometer of the Swiss muon source at Paul Scherrer Institute, Villigen, Switzerland. The samples were mounted on a thin sample holder using diluted GE varnish. Zero-field (ZF) and longitudinal-field (LF) experiments were both carried out down to 300~mK. The $\mu$SR data were analyzed by using the \texttt{musrfit} software package~\cite{suter2012musrfit}.

\begin{figure}[h]
	\centering
	\includegraphics[height=10.5cm,width=6cm]{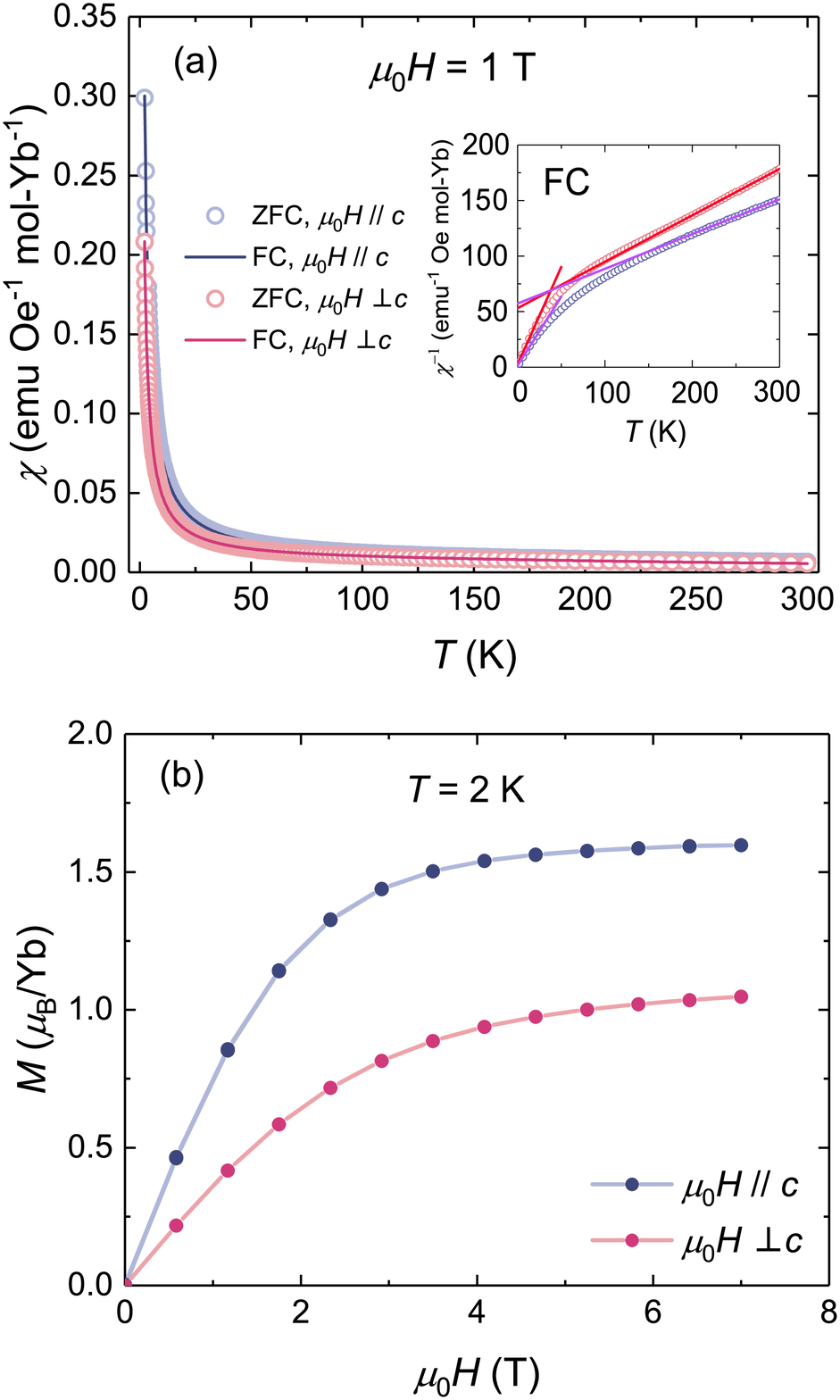}
	\caption{\label{magnetization}(a) Temperature dependence of magnetic susceptibility of Yb(BaBO$_3$)$_3$ measured with ZFC and FC, at $\mu_0H= 1$~T applied parallel and perpendicular to the $c$ axis. The inset shows the inverse of magnetic susceptibility. The lines are Curie-Weiss law fits. (b) Isothermal magnetization at $T=2$~K under applied magnetic fields parallel and perpendicular to the $c$ axis of Yb(BaBO$_3$)$_3$ up to 7 T.}
\end{figure}

\section{RESULTS AND DISCUSSIONS}

\subsection{\label{subsec:sus}Magnetization}

The temperature dependence of the dc magnetic susceptibility $\chi$ at $\mu_0H$ = 1 T down to 2 K is shown in Fig.~\ref{magnetization}(a). Neither a phase transition nor bifurcations between ZFC and FC curves was observed, indicating the lack of a magnetic ordering or spin freezing down to 2 K. In the inset of Fig.~\ref{magnetization}(a), we present the inverse of magnetic susceptibility $\chi^{-1}$ at $\mu_0H$ = 1 T applied parallel and perpendicular to the $c$ axis of single crystal Yb(BaBO$_3$)$_3$. A uniaxial easy axis anisotropy of temperature-dependent susceptibility is observed, $\chi_{H{\parallel}c} > \chi_{H{\perp}c}$. Above 200 K, the magnetic susceptibility data can be well fitted by Curie-Weiss law. $\chi^{-1}$ deviates from a linear dependence on temperature with temperature decreasing below 200 K due to the CEF effect~\cite{guo2019magnetism}. But the Curie-Weiss law is again valid at low temperatures (2-20~K). Such a behavior is also seen in NaBaYb(BO$_3$)$_2$~\cite{guo2019magnetism}, KBaYb(BO$_3$)$_2$~\cite{pan2021specific}, and RbBaYb(BO$_3$)$_2$ \cite{guo2019triangular}, which share a similar chemical composition with Yb(BaBO$_3$)$_3$.\par

The Curie-Weiss fitting in the 200-300 K temperature range yielded the effective magnetic moments for $H \perp c$ and $H \parallel c$ are 4.38 $\mu_{\rm B}$ and 5.06 $\mu_{\rm B}$ respectively, which are comparable to the free Yb$^{3+}$ ion (4$f^{13}$, $^2$F$_{7/2}$, $J = 7/2$, 4.54 $\mu_{\rm B}$). The negative Curie-Weiss temperatures $\theta_{\perp,H}$ = -127.1(2) K and $\theta_{\parallel,H}$ = -182.9(4) K indicate antiferromagnetically coupled Yb$^{3+}$ spins in Yb(BaBO$_3$)$_3$.  The low-temperature Curie-Weiss fitting yields $\theta_{\perp,L}$ = -1.40(7) K and $\theta_{\parallel,L}$ = -1.16(4) K, suggesting small exchange interactions with effective magnetic moments 2.13 $\mu_{\rm B}$ ($H \perp c$) and 2.54 $\mu_{\rm B}$ ($H \parallel c$). The large difference compared with the moment of a free Yb$^{3+}$ ion, was also reported in other Yb-based triangular lattices~\cite{guo2019triangular,baenitz2018naybs,ding2019gapless}.\par

The isothermal magnetization up to 7 T applied parallel and perpendicular to the $c$ axis of the sample at 2 K is shown in Fig.~\ref{magnetization}(b). $M(H)$ curves start to show a non-linear behavior with magnetic field above 2~T for both orientations, and magnetization tends to saturate above 4 T, indicating small exchange interactions in Yb(BaBO$_3$)$_3$. The saturation value of magnetization is 1.60~$\mu_{\rm B}$/Yb and 1.05 $\mu_{\rm B}$/Yb for field along and perpendicular to $c$ axis respectively, which is consistent with the result reported previously \cite{zeng2020nmr}.

\begin{figure}
	\centering
	\includegraphics[height=15.5cm,width=6.5cm]{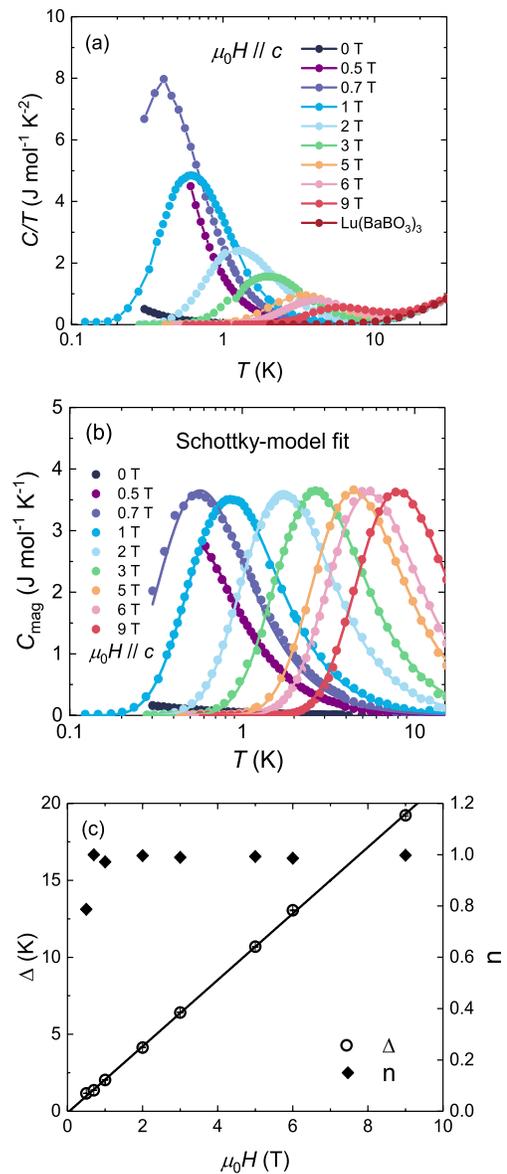}
	\caption{\label{hc}(a) Specific heat of single crystal Yb(BaBO$_3$)$_3$ at different magnetic fields up to 9 T plotted as $C/T$ vs $T$. Specific heat of  nonmagnetic polycrystal Lu(BaBO$_3$)$_3$ is also shown for comparison. (b) Temperature dependence of magnetic heat capacity under different magnetic fields. The colored lines are fitting curves corresponding to Schottky function Eq. (\ref{sch}). (c) Field dependence of the energy gap $\Delta$ and the concentration of Schottky centers $n$ derived from Schottky fitting. The black line is a linear fit.}
\end{figure}

\subsection{\label{subsec:SP}Specific heat}
In Fig.~\ref{hc}(a), we present the specific heat coefficient $C/T$ of Yb(BaBO$_3$)$_3$ at several applied magnetic fields, and $C/T$ of Lu(BaBO$_3$)$_3$ for comparison. No sharp anomaly for magnetic transitions is observed, which is consistent with the magnetization measurements. With magnetic fields applied, the magnetic specific heat $C_{\rm mag}$ of Yb(BaBO$_3$)$_3$ obtained by subtracting the phonon contribution shows a peak, which retains its amplitude but shifts to higher temperatures with increasing applied magnetic fields (Fig. \ref{hc}(b)), indicating a typical Schottky anomaly. The peaks can be fitted using the two-level Schottky function:
\begin{equation}
\label{sch}
C_{\rm mag}=nR{(\frac{\Delta}{T})}^{2}\frac{e^{-\frac{\Delta}{T}}}{{(1+e^{-\frac{\Delta}{T}})}^{2}},
\end{equation}
where $n$ is the concentration of Schottky centers, $R$ is the molar gas constant, and $\Delta$ is the energy separation between two levels. As shown in Fig.~\ref{hc}(c), $n\approx1$ for the applied magnetic fields larger than 0.7~T, indicating the two levels are completely opened when $\mu_0H > 0.7$~T.  The energy gap $\Delta$ shows a linear behavior with applied magnetic fields, reflecting a Zeeman splitting effect. It is worth noting that the magnetic specific heat at zero field cannot be fitted well by Eq.~(\ref{sch}), which will be discussed later. The entropy calculated by integrating $C_{\rm mag}/T$ reaches $R$ln2 for data with magnetic fields larger than 0.7~T, which is consistent with $n\approx1$. While for zero field, the entropy at 0.3 K only reaches 2.5$\%$ of $R$ln2, indicative of a considerable remaining entropy below 0.3~K.\par

In general, hyperfine interactions in lanthanide with $4f$ electrons provide nuclear energy levels, which result in nuclear Schottky anomaly~\cite{tari2003specific}. The nuclear Schottky anomalies at zero field usually occur in the temperature region of 10$^{-2}$ K, and an external applied field on the order of 10~T cannot make a significant difference on the nuclear Schottky effect~\cite{tari2003specific,grivei1995nuclear}. For $C/T$ of Yb(BaBO$_3$)$_3$, we observe remarkable shifts of the Schottky peak with the applied magnetic fields of a few Tesla. Hence the nuclear Schottky can be excluded in accounting for the observed anomaly shown in Fig.~\ref{hc}(b). On the other hand, energy levels due to CEF effect can also lead to a Schottky anomaly. In Yb-based triangular lattices, $4f^{13}$ electrons of the Yb$^{3+}$ ions form four Kramers doublets due to crystal electrical fields~\cite{ross2011quantum,li2016anisotropic,onoda2010quantum}. The energy gap between the ground-state Kramers doublet and the first excited state is on the order of 10 meV (i.e., 100 K)~\cite{li2017crystalline,ding2019gapless,baenitz2018naybs,dai2021spinon}. However, the energy gap of Yb(BaBO$_3$)$_3$ at zero field $\Delta(0)$ derived from the field dependence of the energy gap $\Delta(H)$ is nearly 0, which is particularly far from 100 K. Therefore the low temperature Schottky anomaly is not due to the the energy gap between the ground-state Kramers doublet and the first excited state. \par

We attribute the Schottky anomalies observed in specific heat of Yb(BaBO$_3$)$_3$ to the opening of the ground-state Kramers doublet with magnetic fields applied. At zero field, the increase of $C_{\rm mag}/T$ below 1~K is due to the electronic spin excitations, which may exist in some quantum magnets where exchange interaction is weak, and dipole-dipole interaction is dominant to create an anisotropy~\cite{quilliam2010disorder}. Low-energy spin excitations with a gap resulting from such anisotropy have been reported in Gd$_2$Sn$_2$O$_7$~\cite{quilliam2007evidence}. The magnetic specific heat, similar to the one for Yb(BaBO$_3$)$_3$, is proportional to $(1/{T^2})e^{-\frac{\Delta}{T}}$ at zero field. Spin excitations have also been reported in a dipolar magnet Yb$_3$Ga$_5$O$_{12}$ from the specific heat and inelastic neutron-scattering experiments~\cite{lhotel2021spin}. \par

In Yb(BaBO$_3$)$_3$, both the small Curie-Weiss temperatures determined from low temperature susceptibility and the isothermal magnetization data suggest that the exchange interaction is weak. Considering that $J_{\pm}$ = 0.90 K, $J_{zz}$ = 0.98 K for YbMgGaO$_4$~\cite{li2015rare} and $J_{\pm}$ = 0.18 K, $J_{zz}$ = 0.23 K for NaBaYb(BO$_3$)$_2$~\cite{pan2021specific}, which have a similar structure and comparable distances between Yb$^{3+}$ ions with Yb(BaBO$_3$)$_3$, the exchange interaction in Yb(BaBO$_3$)$_3$ is also likely to be on the order of 0.1~K. Indeed a thermodynamic property study has revealed that the dominant interaction is the long-range dipole-dipole coupling in Yb(BaBO$_3$)$_3$~\cite{bag2021realization}. The dipole-dipole interaction can induce an anisotropy for the magnetic properties~\cite{quilliam2010disorder}. Such anisotropy may affect the degeneration of the ground-state Kramers doublet, creating gapped low-energy spin excitations at zero field and small magnetic fields (less than 0.7 T in the case of Yb(BaBO$_3$)$_3$). \par

\subsection{\label{subsec:musr}$\mu$SR}

\begin{figure}
	\centering
	\includegraphics[height=11cm,width=7cm]{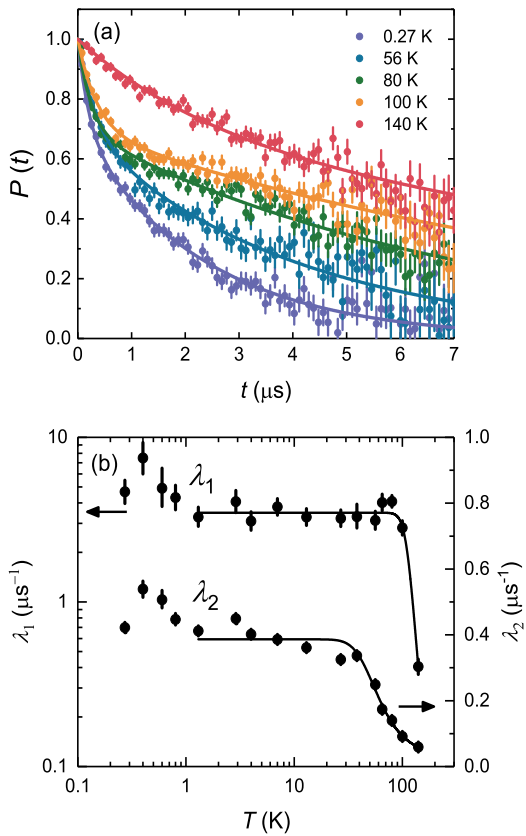}
	\caption{\label{zfmusr} (a) The ZF-$\mu$SR spectra of Yb(BaBO$_3$)$_3$ at different temperatures. The solid lines correspond to the fitting function of Eq. (\ref{asy}). (b) Temperature dependence of muon relaxation rates $\lambda_1$ and $\lambda_2$ obtained from fitting. The solid lines are results of fitting by means of the Orbach function Eq. (\ref{orbach}).}
\end{figure}

As muon is extremely sensitive to small local magnetic fields, we performed ZF-$\mu$SR measurements on Yb(BaBO$_3$)$_3$ down to 270 mK. ZF-$\mu$SR spectra at several representative temperatures are shown in Fig.~\ref{zfmusr}(a). Neither oscillations nor an initial asymmetry loss was observed throughout the whole temperature range, confirming the absence of long-range magnetic order or spin-freezing \cite{zheng2005coexistence}. Furthermore, the lack of polarization recovery to 1/3 of the initial asymmetry value rules out a static random field distribution \cite{uemura1985muon}, demonstrating the dynamic nature of magnetism in  Yb(BaBO$_3$)$_3$. \par

The normalized ZF-$\mu$SR spectra after subtracting the background signal can be well fitted by the function:
\begin{equation}
\label{asy}
P(t)=f_{1}e^{-{{\lambda}_{1}}t}+(1-f_{1})e^{-{{\lambda}_{2}}t}
\end{equation}
where $\lambda_1$ and $\lambda_2$ are muon spin relaxation rates, $f_1$ is the fraction of the first exponential component. The value of $f_1$ was found to be temperature-independent, therefore it was fixed at its average value 0.3. We attribute the two exponential components with their temperature-independent fractions to two muon sites in Yb(BaBO$_3$)$_3$, which will be discussed later.
\par

The temperature dependence of relaxation rates $\lambda_1$ and $\lambda_2$ is shown in Fig.~\ref{zfmusr}(b). $\lambda_1$ is significantly larger than $\lambda_2$, while they show similar temperature dependence and both have a plateau-like behavior in the intermediate temperature range. $\lambda_2$ remains almost temperature-independent for $1 \le T \le 40$ K. Above 40 K, $\lambda_2$ decreases gradually as the temperature increases. $\lambda_1$ remains independent of temperature over a wider temperature range, at least up to 100 K, then a drop of $\lambda_1$ was found at 140 K. Muon diffusion can lead to such a behavior, however, both $\lambda_1$ and $\lambda_2$ should drop at the same temperature in this case. The temperature dependence of $\lambda_{1,2}$ above 1 K can be described by an Orbach process \cite{de2003absence}. Orbach process is a two-phonon scattering process with an excited CEF level, which was generally observed in some insulating oxides \cite{orbach1961spin,arh2021ising}. A global fit ($1 \le T \le 140$ K) was performed for $\lambda_1$ and $\lambda_2$ with the form:
\begin{equation}
	\label{orbach}
	\lambda^{-1}={\lambda_{0}}^{-1}+{B_{\rm me}}\exp({-\Delta_{\rm CEF}}/T)
\end{equation}
where $\lambda$ denotes the relaxation rate $\lambda_1$ or $\lambda_2$, $\lambda_{0}$ is the saturation value of the relaxation rate, $B_{\rm me}$ models the magnetoelastic coupling of the Yb$^{3+}$ electronic spin with the phonon bath. $\Delta_{\rm CEF}$ is the energy difference between CEF levels involved in Orbach process. We get $\Delta_{\rm CEF,2}$ = 17.2(8) meV from the fit of temperature dependence of $\lambda_2$. $\Delta_{\rm CEF,2} \approx$ 170 K (in $k_BT$) is consistent with the temperature where $\chi^{-1}$ deviates from a Curie-Weiss dependence due to CEF, and comparable with the CEF energy gap between the ground state doublet and the first excited state of other Yb-based triangular lattices~\cite{li2017crystalline,ding2019gapless,baenitz2018naybs,dai2021spinon}.  Because there are too few data in the decline range of $\lambda_1$, we can only get qualitative estimation. The fit using Eq.~(\ref{orbach}) yields $\Delta_{\rm CEF,1}$ = 107(17) meV, which may not only be the energy difference of the ground-state doublet and the first excited state, higher excited states may also be associated.\par

As shown in Fig.~\ref{structure}(b), Rietveld refinement of XRD on Yb(BaBO$_3$)$_3$ has revealed that Yb$^{3+}$ ions are located at two different sites~\cite{gao2018synthesis}. The surrounding charge distributions of the two distinct Yb$^{3+}$ ions are different, thus resulting in two different CEFs. Different CEF energy gaps have also been observed in a Yb-based pyrochlore Yb$_2$Ti$_2$O$_7$ with two Yb$^{3+}$ sites \cite{gaudet2015neutron}, while the difference between the CEF energy gaps is less significant. For Yb(BaBO$_3$)$_3$, a sketch of the triangular lattice with two inequivalent Yb$^{3+}$ ions and two different muon sites is shown in Fig.~\ref{structure}(b). Since $\mu$SR is a local probe on an atomic scale, $\mu1$ and $\mu2$, which are close to Yb1 and Yb2, respectively, experience different CEFs. The ratio of the occupation number of two Yb$^{3+}$ sites is $1:2$, which is almost consistent with the ratio of two muon sites $3:7$ in Eq.~(\ref{asy}), indicating that different CEFs do exist in Yb(BaBO$_3$)$_3$.  \par

Below $T=1$~K, both $\lambda_1$ and $\lambda_2$ exhibit a slightly upward trend (Fig.~\ref{zfmusr}(b)).  Although the detail behavior of $\lambda_{1,2}(T)$ at low temperatures can not be determined due to the limited number of points, the deviation of plateau-like behavior below 1~K can be seen clearly.  We note that the specific heat $C/T$ at zero field also starts to increase below 1~K with temperature decreasing. The anomaly of $\lambda_{1,2}$ and $C/T$ share the same temperature range, indicating they are induced by the same physics. Spin excitation has been suggested to account for the anomaly in $C/T$ as discussed in Sec.~\ref{subsec:SP}, and it may also lead to an increase of muon-spin relaxation rates~\cite{yaouanc2011muon}. \par

\begin{figure}
	\centering
	\includegraphics[height=8cm,width=8.5cm]{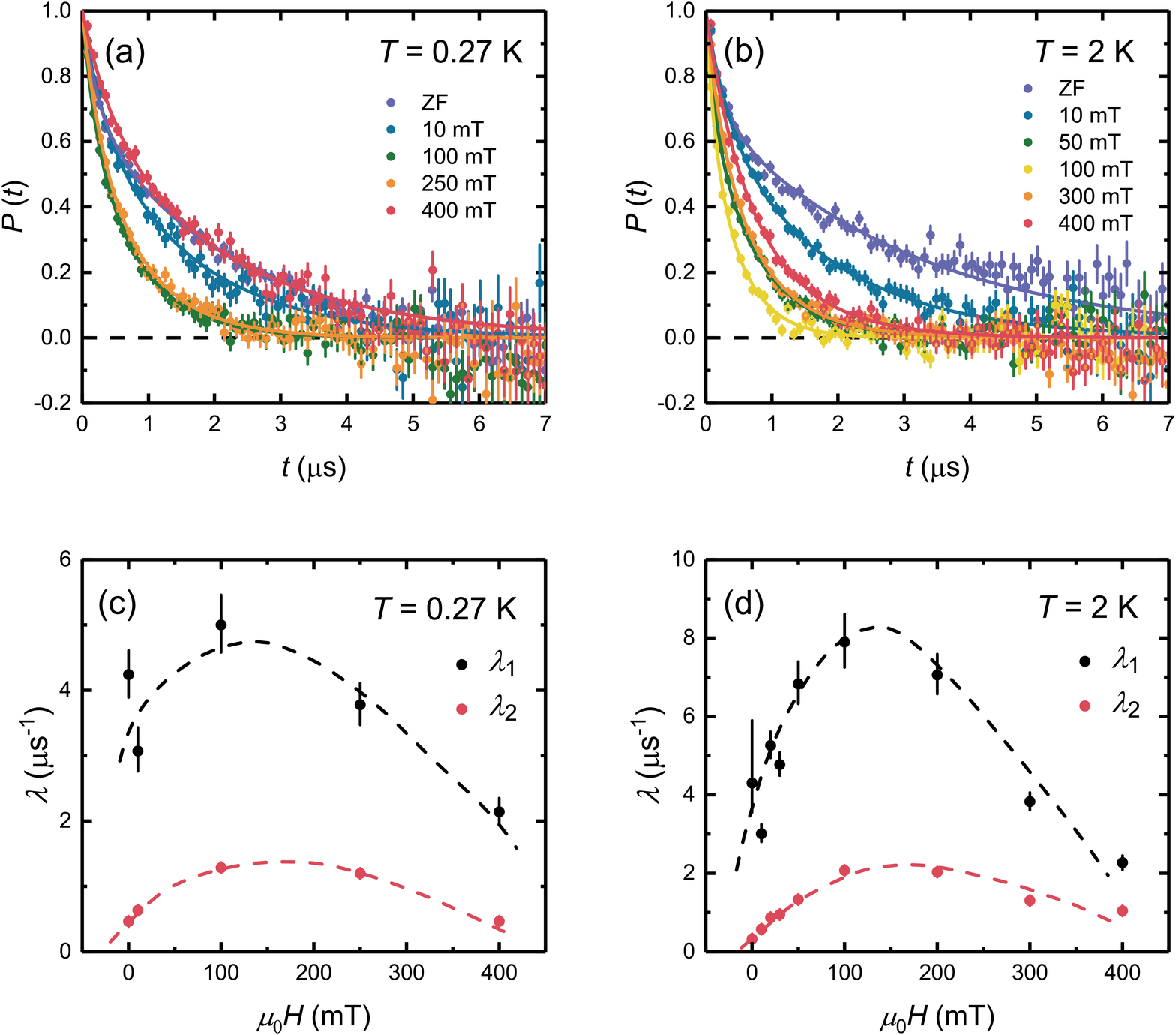}
	\caption{\label{lfmusr} LF-$\mu$SR spectra of Yb(BaBO$_3$)$_3$ under applied fields up to 400 mT at (a) 0.27 K and (b) 2 K. No decouple phenomenon is observed. Field dependence of two muon relaxation rates $\lambda_1$ and $\lambda_2$ at 0.27 K (c) and 2 K (d). $\lambda_1$ and $\lambda_2$ both reach their maximum at about 150~mT instead of being quenched monotonically by magnetic field. The dashed lines are a guide to the eye. }
\end{figure}

To further investigate the spin dynamics, LF-$\mu$SR measurements were performed at $T=270$~mK and 2 K. As seen in Fig.~\ref{lfmusr}(a)-(b), the relaxation persists also in an applied magnetic field. Even a field of 400 mT is insufficient to decouple the muon spins, suggesting that the magnetic field at muon stopping site is dynamic. The LF-$\mu$SR spectra are also well fitted by function Eq.~(\ref{asy}), whereas $\lambda_1$ and $\lambda_2$ here are muon spin relaxation rates with longitudinal fields applied. $f_1$ is 0.3, which is consistent with the ZF value. \par

In Fig.~\ref{lfmusr}(c)-(d), we present the field dependence of $\lambda_1$ and $\lambda_2$ at 270~mK and 2 K, respectively. The field dependences of the two relaxation rates are similar but quite unusual. Generally, the magnetic field suppresses the dynamical relaxation rate \cite{yaouanc2011muon,li2016muon}. However, here both relaxation rates show an unexpected initial increase instead of being quenched monotonically with magnetic field. A maximum was observed at 150~mT, after which two relaxation rates start to decrease with increasing the magnetic filed. The field-induced hump in the muon spin relaxation rates can be due to the level-cross resonance (LCR) between the muon Zeeman splitting and the induced quadrupole coupling of the muon's nearest-neighbor nuclei~\cite{kreitzman1986longitudinal}. The LCR occurs at a field of $B\approx\omega_Q/\gamma_{\mu}$, where $\omega_Q$ is the transition frequencies between the nuclear quadrupole energy levels and $\gamma_{\mu}$ is the muon's gyromagnetic ratio. However, The transition frequency $\omega_Q$ is on the order of 1$\sim$10 MHz, which means the LCR occurs at a magnetic field of 1$\sim$10 mT~\cite{kreitzman1986longitudinal,lindon2016encyclopedia,kundu2020signatures,kuroiwa2006possible}. The field maximum of muon spin relaxation rates  in Yb(BaBO$_3$)$_3$ is observed only around 150 mT, an order of magnitude larger than the field where LCR occurs. Thus, the LCR effect can be excluded.\par 

We attribute the initial increase of $\lambda_1$ and $\lambda_2$ to a field-induced increase of the density of spin excitations, which competes with the quenching effect of the applied magnetic field. A maximum of the relaxation rates at 150~mT is eventually formed due to such competition. A similar behavior of field dependence of the muon spin relaxation rate was reported in frustrated pyrochlore Tb$_2$Sn$_2$O$_7$, where the relaxation rate reached a maximum at about 50~mT at low temperatures~\cite{de2006spin}. Since we find a maximum at higher magnetic field, its effect on spin excitations is stronger than that of Tb$_2$Sn$_2$O$_7$.

\section{CONCLUSIONS}
In conclusion, we performed magnetization, specific heat and $\mu$SR measurements on single crystals of Yb(BaBO$_3$)$_3$, a disorder-free Yb triangular lattice compound. The absence of long-range antiferromagnetic order and spin freezing are confirmed down to $T=0.27$~K. A two-level Schottky anomaly due to the opening of the ground-state Kramers doublet with applied magnetic fields $\mu_0H >0.7$~T is observed from the low-temperature specific-heat measurements. At zero field, the increase of $C_{\rm mag}/T$ below 1~K is due to the electronic spin excitations, which often exist in quantum magnets where dipole-dipole interaction is dominant to create an anisotropy of magnetic properties, and affect the degeneration of the ground-state doublet. This insight is supported by the $\mu$SR experiments which reveal dynamic magnetic fluctuations at low temperatures, i.e., the increase of muon spin relaxation rate $\lambda$ below 1 K, and the unusual field dependence of relaxation rate $\lambda(H)$.\par

Our work shows the interesting behavior of a quantum dipolar magnet, while providing a clear example where dipolar interaction is dominant in a frustrated geometry. However, more experiments and theoretical works are needed to shed light on whether frustration exists and plays an important role for the low temperature behaviors of Yb(BaBO$_3$)$_3$. 

\begin{acknowledgments}
This work is based on experiments performed at the Swiss Muon Source S$\mu$S, Paul Scherrer Institute, Villigen, Switzerland. The research performed in this work was supported by the National Natural Science Foundations of China, Nos.~12034004, 12174065 and 11874158, the Shanghai Municipal Science and Technology (Major Project Grant No.~2019SHZDZX01 and No.~20ZR1405300), and the National Key Research and Development Program of China, No.~2019YFE0100400.
\end{acknowledgments}

%

\end{document}